\documentclass[10pt]{article}
\usepackage{appendix}
\usepackage{epsfig}
\usepackage{capt-of}

\usepackage[activeacute,english]{babel}
\usepackage{latexsym}

\newcommand{\be}{\begin{equation}}
\newcommand{\ee}{\end{equation}}
\newcommand{\bea}{\begin{eqnarray}}
\newcommand{\eea}{\end{eqnarray}}

\newcommand{\gapp}{\mathrel{\raise.3ex\hbox{$>$}\mkern-14mu
              \lower0.6ex\hbox{$\sim$}}}
\newcommand{\lapp}{\mathrel{\raise.3ex\hbox{$<$}\mkern-14mu
              \lower0.6ex\hbox{$\sim$}}}

\newcommand{\lsim}{\mbox{\raisebox{-.9ex}{~$\stackrel{\mbox{$<$}}{\sim}$~}}}

\renewcommand\({\left(}
\renewcommand\){\right)}
\renewcommand\[{\left[}
\renewcommand\]{\right]}

\newcommand\eq[1]{Eq.~(\ref{#1})}

\def\calb{{\cal B}}

\def\calp{{\cal P}}

\def\calt{{\cal T}}

\def\calpz{{\calp_\zeta}}
\def\calbz{{\calb_\zeta}}

\newcommand\bfA{{\mathbf A}}

\newcommand\bfd{{\mathbf d}}

\newcommand\bfk{{\mathbf k}}

\newcommand\bfn{{\mathbf N}}

\newcommand\bfp{{\mathbf p}}

\newcommand\bfx{{\mathbf x}}

\newcommand\sub[1]{_{\rm #1}}
\newcommand\su[1]{^{\rm #1}}

\newcommand\half{^{1/2}}

\newcommand\threehalf{^{3/2}}

\newcommand{\fnl}{f\sub{NL}}

\newcommand{\tnl}{\tau\sub{NL}}
\newcommand{\gnl}{g\sub{NL}}

\newcommand\pz{P_\zeta}
\newcommand\bz{B_\zeta}

\newcommand\gz{g_\zeta}
\newcommand{\no}{\nonumber}

\begin{document}

\title{\begin{flushright}
\normalsize PI/UAN-2009-352FT
\end{flushright}
\vspace{5mm}
{\bf Non-gaussianity at tree and one-loop levels from vector field perturbations}}

\author{
\textbf{C\'esar A. Valenzuela-Toledo$^{1,}$\thanks{e-mail: \texttt{cavalto@ciencias.uis.edu.co}}}, \textbf{Yeinzon Rodr\'{\i}guez$^{1,2,}$\thanks{e-mail: \texttt{yeinzon.rodriguez@uan.edu.co}}}, \textbf{David H. Lyth$^{3,}$\thanks{e-mail: \texttt{d.lyth@lancaster.ac.uk}}} \\ \\
\textit{$^1$Escuela de F\'{\i}sica, Universidad Industrial de Santander,}  \\
\textit{Ciudad Universitaria, Bucaramanga, Colombia} \\
\textit{$^2$Centro de Investigaciones, Universidad Antonio Nari\~no,}\\
\textit{Cra 3 Este \# 47A-15, Bogot\'a D.C., Colombia}\\
\textit{$^3$Department of Physics, Lancaster University,}  \\
\textit{Lancaster LA1 4YB, UK} \\
}

\maketitle

\begin{abstract}
\noindent We study the spectrum $\calp_\zeta$ and bispectrum $B_\zeta$ of the primordial curvature perturbation $\zeta$ when the latter is generated by
scalar and vector field perturbations.
The tree-level and one-loop contributions from vector field perturbations are worked out considering the possibility that
the one-loop contributions may be dominant over the tree-level terms (both (either) in $\calp_\zeta$
and (or) in $B_\zeta$) and viceversa. The level of non-gaussianity in the bispectrum, $f_{NL}$,
is calculated and related to  the level of statistical anisotropy in the power spectrum, $g_\zeta$. 
For very small amounts of statistical anisotropy in the power spectrum, the level of 
non-gaussianity may be very high, in some cases exceeding the current observational limit.
% and therefore observable.

\end{abstract}

\section{Introduction}
The anisotropies in the temperature of the cosmic microwave background (CMB) radiation, which have strong connections with the origin of the
large-scale structure in the observable Universe, is one of hottest topics in modern cosmology. The properties of the CMB temperature anisotropies
are described in terms of the spectral functions, like the spectrum, bispectrum, trispectrum, etc., of the primordial curvature perturbation $\zeta$
\cite{cogollo}.
In most of the cosmological models the $n$-point correlators of $\zeta$ are supposed to be translationally and rotationally invariants. However, violations
of such invariances entail modifications of the usual definitions for the spectral functions in terms of the statistical descriptors
\cite{acw,carroll,armendariz}.
These violations may be consequences either of the presence of vector field perturbations
\cite{armendariz,vc,vc2,RA2,ys,gmv,kksy,gmv2,gvnm,koh,himmetoglu,himmetoglu2,himmetoglu3,himmetoglu4,dklr,dkl,dkw,bdmr,dkw2},
spinor field perturbations \cite{bohmer,shan}, or p-form perturbations \cite{germani,kobayashi,
koivisto,germani2,koivisto2}, contributing significantly to $\zeta$,
of anisotropic expansion \cite{kksy,himmetoglu4,bohmer,koivisto,ppu1,gcp,ppu2,watanabe,bamba,dechant} or of an
inhomogeneous background \cite{carroll,armendariz,dklr}.
Violation of the statistical isotropy (i.e., violation of the rotational invariance in the
$n$-point correlators of $\zeta$) seems to be present in the data \cite{ge2,ge,app,hl,samal} and, although its statistical
significance is still low, the continuous presence of anomalies in every CMB data analysis (see for instance Refs.
\cite{hou,hoftuft,hansen,dipole1,dipole2,dipole3,dvorkin,land1,land2,oliveira,schwarz,tegmark,bunn1}) suggests the evidence might be decisive in the forthcoming years. Since the
statistical anisotropy is observationally low, it entails a big problem when vector fields are present during inflation, because they
generically lead to a high amount of statistical anisotropy, higher than that coming from observations
\cite{gmv,kksy,dklr}. To solve this problem, people use different mechanisms in order to make those models consistent with observation,
for example, using a triad of orthogonal vectors \cite{armendariz,bento}, a large number of
identical randomly oriented vectors fields \cite{gmv}, or assuming that the contribution of
vector fields to the total energy density is negligible \cite{kksy,dklr}.

Assuming statistical homogeneity of the curvature perturbation (i.e., translational invariance
of the $n$-point correlators of $\zeta$) and pretending that during
inflation the rotational invariance is broken by the
presence of a vector field that points along the direction of some unit vector $\hat{\bfd}$,
the most general form of the power spectrum changes from
$P(k)$ to $P'(\bfk)$: \cite{acw}
\be\label{statan}
P'(\bfk)=P(k)(1+\gz\;(\hat{\bfd}\cdot\hat{\bfk})^2+\ldots) \,,
\ee
where $\hat{\bfk}$ is the unit vector along the direction of the wave vector $\bfk$, and
$g_\zeta$ is the level of statistical anisotropy. The above formula gives us the
primordial power spectrum that takes into account the leading effects of violations of statistical isotropy by
the presence of some vector field in the inflationary era.
Taking all uncertainties into account, observation is consistent with violations of the 
statistical isotropy at the level of $30\%$. Recent data
analysis \cite{ge2,ge,app,hl,samal} suggest the existence of violations of statistical isotropy in the five-year data from the NASA's WMAP satellite
\cite{wmap}. A recent study \cite{ge2} of the CMB temperature perturbation finds weak evidence
for statistical anisotropy; they keep only the
leading (quadrupolar) term of \eq{statan}:
\be
\calpz(\bfk) = \calp_\zeta\su{iso}(k) \( 1 + g_\zeta
(\hat{\bfd}\cdot \hat \bfk)^2 \) \,,
\label{curvquad} \ee
and find %$g_\zeta \simeq 0.15\pm0.04$ 
$g_\zeta \simeq 0.290\pm0.031$ at the 68\% confidence level with $\hat{\bfd}$ in a
specified direction.
The authors point out though that systematic uncertainties could  make $g_\zeta$
compatible with zero. A related work \cite{pullen} shows that the lowest detectable
value for $|g_\zeta|$ from the expected performance of the NASA's WMAP satellite \cite{wmap} (currently in operation) is $|g_\zeta| \simeq 0.1$.
The same analysis gives the lowest detectable value from the expected
performance of the ESA's PLANCK satellite \cite{planck}: $|g_\zeta| \simeq 0.02$.

As the $\gz$ parameter has observational bounds and works, together with the non-gaussianity
parameters $\fnl$, $\tnl$, $\gnl$, etc., as
statistical descriptors for $\zeta$, it could be a crucial tool to discriminate between some of the most usual cosmological models. In a recent
paper \cite{dkl}, the authors point out the possibility that a vector field causes part
of the primordial curvature perturbation and show that the
non-gaussianity parameter $\fnl$ is statistically anisotropic, being in principle observable.
They also found a consistency relation between the
parameters $\gz$ and $\fnl$. In such a work the authors included both vector and scalar field
perturbations and kept only the lowest order terms
in the expressions for the bispectrum $\bz$ and spectrum $\pz$. As a crucial point, they
assumed that the contributions to the spectrum from vector field
perturbations were smaller than those coming from scalar fields and in an opposite way for
the bispectrum.

In this paper we explore other possibilities than that explored in Ref. \cite{dkl}, extending
the analysis to include higher order
contributions and studying the possibility that the loop contributions may dominate over the
tree-level terms\footnote{For a similar study regarding the trispectrum $T_\zeta$, see Ref. \cite{newvr}.}. We begin our study by giving some useful
formulas and calculating the 1-loop contribution to the spectrum $\calp_\zeta$ and bispectrum $\bz$ including vector and
scalar fields perturbations. We then calculate the
level of non-gaussianity in the bispectrum including the loop contributions and write down
formulas that relate $\fnl$ and $\gz$. Finally, comparison with the current observational
bounds for $\fnl$ and $\gz$ is done.

%%%%%%%%%%%%%%%%%%%%%%%%%%%%%%%%%%%%%%%%%%%%%%%%%%%%%%%%%%%%%%%%%%%%%%%%%%
\section{Spectrum and bispectrum from vector field perturbations}		%%
%%%%%%%%%%%%%%%%%%%%%%%%%%%%%%%%%%%%%%%%%%%%%%%%%%%%%%%%%%%%%%%%%%%%%%%%%%

In a recent paper \cite{dklr} the $\delta N$ formalism \cite{starobinsky2,ss,tanaka,sasaki1,lr1} was extended to
include the possible statistical anisotropy in primordial curvature perturbation $\zeta$
originated from vector field
perturbations. It was shown in that paper that the curvature perturbation, in the simplest case where $\zeta$ is generated by one scalar field and
one vector field and assuming that the anisotropy in the expansion of the Universe is negligible, can be calculated
up to quadratic terms by means of the following truncated expansion:\footnote{This expression
corrects Eq. (3.14) of Ref. \cite{dklr}, and Eq. (3) of Ref. \cite{dkl}, where a factor 2 in
the fourth term of the expansion is missing.}
\be
\zeta(\bfx)\equiv\delta N (\phi(\bfx),A_i(\bfx),t)=N_\phi \delta\phi + N_A^i\delta A_i+\frac{1}{2}N_{\phi\phi}(\delta\phi)^2+
N_{\phi A}^i\delta\phi\delta A_i+\frac{1}{2}N_{AA}^{ij}\delta A_i \delta A_j \,,\label{deltan}
\ee
where 
\be
N_\phi\equiv\frac{\partial N}{\partial \phi}\,,\quad
N_A^{i}\equiv\frac{\partial N}{\partial A_i}\,,\quad
N_{\phi\phi} \equiv\frac{\partial^2 N}{\partial \phi^2}\,,\quad
N_{AA}^{ij}\equiv\frac{\partial^2 N}{\partial A_i\partial A_j}\,,
\quad N_{\phi A}^i\equiv\frac{\partial^2 N}{\partial A_i\partial\phi}\,,
\ee
$\phi$ being the scalar field and ${\bf A}$ the vector field, with $i$ denoting the spatial
indices running from 1 to 3. Now, we define the power spectrum $\pz$ and the
bispectrum $\bz$ for the primordial curvature perturbation, through the Fourier modes of $\zeta$, as:
\bea
\langle\zeta(\bfk)\zeta(\bfk')\rangle&\equiv&(2\pi)^3\delta(\bfk+\bfk')P_\zeta(\bfk)
\;\;\equiv \;\;(2\pi)^3\delta(\bfk+\bfk')\frac{2\pi^2}{k^3}\calpz(\bfk) \,,
\label{spdef}\\
\langle\zeta(\bfk)\zeta(\bfk')\zeta(\bfk'')\rangle&\equiv&(2\pi)^3\delta(\bfk+\bfk'+\bfk'')\bz(\bfk,\bfk',\bfk'')
\;\; \equiv \;\;(2\pi)^3\delta(\bfk+\bfk'+\bfk'')\frac{4\pi^4}{k^3k'^3}\calbz(\bfk,\bfk',\bfk'') \,.
\label{bsdef}
\eea
Using \eq{deltan} and the definitions given in Eqs. (\ref{spdef}) and (\ref{bsdef}), it was
found in Ref. \cite{dklr} that
the tree-level contribution to the spectrum is of the form shown in \eq{curvquad}.
In addition, an analogous form for
the contribution to $\fnl$ was given in Ref. \cite{dkl}, showing that both, $\pz$ and
$\fnl$ have anisotropic contributions coming from the vector field perturbation. The one-loop
correction to the spectrum was
also given in Ref. \cite{dklr}; however they kept it in an integral form. In this paper we
give the loop contribution to
the bispectrum, and also estimate the integrals  in order to get an order of magnitude for
$\fnl$. To estimate the
integrals we follow a similar procedure as that presented in Refs. \cite{lyth0,lyth2,lyth1},
but this time including also
the vector field. The expressions for $\calpz$ and $\calbz$, defined in Eqs.
(\ref{spdef}) and (\ref{bsdef}) and considering contributions up to one-loop order, are:\footnote{
Eq. (\ref{Pzetal}) corrects a mistake in Eq. (4.12) of Ref. \cite{dklr} where the infinitesimal
volume element $d^3p$ was incorrectly expressed in terms of $dp$.}
\bea
\calp_\zeta^{\rm tree}(\bfk) &=& N_\phi^2 \calp_{\delta \phi}(k) +
N_A^iN_A^j\calt_{ij}(\bfk)  \nonumber\\
&=&
 N_\phi^2 \calp_{\delta \phi}(k) +
N_A^2 \calp_+(k) + ({\bf N}_A\cdot\hat\bfk )^2 \calp_+(k)  \( r\sub{long} - 1  \)\label{Pzetat}
\,, \eea
\bea
\calp_\zeta^{\rm 1-loop} (\bfk) &=& \int
\frac{d^3p k^3}{4\pi|\bfk + \bfp|^3 p^3}
\[\frac{1}{2} N_{\phi \phi}^2  \calp_{\delta \phi} (|\bfk + \bfp|)
\calp_{\delta \phi}(p) +
N_{\phi A}^i N_{\phi A}^j \calp_{\delta \phi} (|\bfk + \bfp|)
\calt_{ij}(\bfp)\right. \nonumber \\&&\left. + \frac{1}{2} N_{AA}^{ij} N_{AA}^{kl}\calt_{ik}(\bfk+\bfp)\calt_{jl}(\bfp) \] \,, \label{Pzetal}
\eea
\bea
\calb_\zeta^{\rm tree} (\bfk,\bfk',\bfk'') &=& N_\phi^2 N_{\phi \phi}
[\calp_{\delta \phi} (k) \calp_{\delta \phi} (k') + {\rm cyc. \ perm.}]  + N_A^i N_A^k N_{AA}^{mn}
\Big[ \calt_{im}(\bfk)\calt_{kn}(\bfk') + {\rm cyc. \ perm.}\Big] \no\\
 &+& N_\phi N_A^i N_{\phi A}^j \Big[\calp_{\delta \phi} (k)
\calt_{ij}(\bfk') + 5 \ {\rm perm.} \Big] \,, \label{bzt}
\eea
\bea
\calb_\zeta^{\rm 1-loop}(\bfk,\bfk',\bfk'')&=&N_{\phi \phi}^6\int\frac{d^3pk^3k'^3}{4\pi p^3|\bfk+\bfp|^3|\bfk'-\bfp|^3}\calp_{\delta\phi}(p)\calp_{\delta\phi} (|\bfk+\bfp|)
\calp_{\delta \phi}(|\bfk'-\bfp|)\no\\
&+& N_{AA}^{ij}N_{AA}^{kl}N_{AA}^{mn}\int\frac{d^3pk^3k'^3}{4\pi p^3|\bfk+\bfp|^3|\bfk'-\bfp|^3}\calt_{il}(\bfp)\calt_{kn}(\bfk+\bfp)\calt_{jm}(\bfk''-\bfp)\no\\
&+&N_{\phi \phi}N_{\phi A}^{i}N_{\phi A}^{j}\int\frac{d^3pk^3k'^3}{4\pi p^3|\bfk''+\bfp|^3|\bfk'-\bfp|^3}\bigg\{\calp_{\delta\phi}(p)\calp_{\delta\phi} (|\bfk''+\bfp|)\calt_{ij}(\bfk'-\bfp)\no\\
&+&\calp_{\delta\phi}(p)\calp_{\delta\phi}(|\bfk'-\bfp|)\calt_{ij}(\bfk''+\bfp)+\calp_{\delta\phi}(|
\bfk'-\bfp|)\calp_{\delta\phi}(|\bfk''+\bfp|)\calt_{ij}(\bfp)\bigg\}\no\\
&+&N_{\phi A}^{i}N_{\phi A}^{j}N_{AA}^{kl}\int\frac{d^3pk^3k'^3}{4\pi p^3|\bfk''+\bfp|^3|\bfk'-\bfp|
^3}\bigg\{\calp_{\delta\phi}(p)\calt_{ik}(\bfk'-\bfp)\calt_{jl} (\bfk''+\bfp)\no\\
&+&\calp_{\delta\phi}(|\bfk''+\bfp|)\calt_{ik}(\bfp)\calt_{jl}(\bfk'-\bfp)+\calp_{\delta\phi}(|
\bfk'-\bfp|)\calt_{ik}(\bfp)\calt_{jl}(\bfk''+\bfp)\bigg\}\label{bsl} \,,
\eea
where 
\be\label{def1}
\calt_{ij}(\bfk)\equiv T_{ij}^{\rm even}(\bfk)\calp_+(k)+iT_{ij}^{\rm odd}(\bfk)\calp_-(k)+T\su{long}_{ij}(\bfk)\calp_{\rm long}(k) \,,
\ee
and
\be
T_{ij}\su{even} (\bfk) \equiv \delta_{ij} -  \hat k_i \hat k_j \,,\qquad
T\su{odd}_{ij} (\bfk) \equiv \epsilon_{ijk}\hat k_k \,,\qquad
T\su{long}_{ij} (\bfk) \equiv \hat k_i \hat k_j\,.
\ee
\eq{Pzetat} was written in the form of \eq{curvquad} with $\hat{\bfd}=\hat{\bfn}_A$, $\bfn_A$
being a vector 
with magnitude  $N_A\equiv\sqrt{N^i_AN^i_A}$, and $r\sub{long}\equiv\calp\sub{long}/\calp_+$, where $\calp\sub{long}$ is the 
power spectrum for the longitudinal component, and $\calp_+$ and $\calp_-$ are the parity conserving and violating
power spectra defined by
\be
\calp_{\pm}\equiv\frac{1}{2}\(\calp_R\pm\calp_L\) \,,
\ee 
with $\calp_R$ and $\calp_L$ denoting the power spectra for the transverse components with right-handed and 
left-handed polarisations \cite{dklr}.

The above expressions can be further separated into different terms: one due to perturbations in the scalar
field, another due to the vector field perturbations, and the other due to the mixed terms:
\bea
\calpz^{\rm tree}(\bfk)&=&\calpz_\phi^{\rm tree}(k)+\calpz_ A^{\rm tree}(\bfk)\label{sst},\\
\calp_\zeta^{\rm 1-loop}(\bfk)&=&\calpz_{\phi}^{\rm 1-loop}(k)+\calpz_{A}^{\rm 1-loop}(\bfk)+\calpz_{\phi A}^{\rm
1-loop}(\bfk)\label{ssl} \,,
\eea
\bea
\calb_\zeta^{\rm tree} (\bfk,\bfk',\bfk'') &=& \calbz_\phi^{\rm tree} (\bfk,\bfk',\bfk'')+\calbz_A^{\rm tree}
(\bfk,\bfk',\bfk'')+\calbz_{\phi A}^{\rm tree} (\bfk,\bfk',\bfk'')\label{bsst} \,,\\
\calb_\zeta^{\rm 1-loop}(\bfk,\bfk',\bfk'')&=&\calbz_\phi^{\rm 1-loop} (\bfk,\bfk',\bfk'')+\calbz_A^{\rm 1-loop}
(\bfk,\bfk',\bfk'')+\calbz_{\phi A}^{\rm 1-loop} (\bfk,\bfk',\bfk'')\label{bssl}\,.
\eea

Observational analysis tell us that the statistical anisotropy in the CMB temperature perturbation
could be observable in a future through current experiments like WMAP or PLANCK. \eq{curvquad} combined with recent studies
\cite{ge2} tells us that the level of statistical anisotropies $g_\zeta$ has an
upper bound and in the best case (99\% confidence level) this is %$\gz\lsim 0.27$ \cite{ge}.
$\gz\lsim 0.383$ \cite{ge2}. During our analysis  we will adopt an
upper bound for $g_\zeta$: $\gz\lsim 0.1$. In order to satisfy the latter observational constraint over the spectrum, we
must be sure that the contributions coming from vector fields in Eqs. (\ref{Pzetat}) and (\ref{Pzetal}) are
smaller than those coming from scalar fields. That means that the first term in  \eq{sst} dominates over all
the other terms, even those coming from one-loop contributions. With the previous conclusion in mind we feel free to make
assumptions over the other contributions, specially for those coming from vector field perturbations.

\section{Vector field contributions to the statistical descriptors}

As we explain in the previous section, our unique restriction from observation is related to
 the amount of statistical
anisotropy present in the spectrum, so we need to be sure that the first term in \eq{sst} always dominates.
In our study we will assume that the terms coming only from the vector field dominate over those coming from
the mixed terms and from the scalar fields only, except for the case of the tree-level
spectrum\footnote{For an actual realisation of
this scenario, we need to show that such constraints are fully satisfied.}. Based on the assumption made, Eqs.
(\ref{sst}) - (\ref{bssl}) take the form:
\bea
\calpz^{\rm tree}(\bfk)&=&\calpz_\phi^{\rm tree}(k)+\calpz_ A^{\rm tree}(\bfk)\label{sst1} \,,\\
\calp_\zeta^{\rm 1-loop}(\bfk)&=&\calpz_{A}^{\rm 1-loop}(\bfk)\label{ssl1} \,,\\
\calb_\zeta^{\rm tree} (\bfk,\bfk',\bfk'') &=&\calbz_A^{\rm tree} (\bfk,\bfk',\bfk'')\label{bsst1} \,,\\
\calb_\zeta^{\rm 1-loop}(\bfk,\bfk',\bfk'')&=&\calbz_A^{\rm 1-loop} (\bfk,\bfk',\bfk'')\label{bssl1} \,.
\eea
The above expressions lead us to four different ways that allow us to study and probably get a high
level of non-
gaussianity\footnote{Our assumption is inspired in the one given in Ref. \cite{lyth2}.
In that work the authors use two scalar fields instead of one scalar and one vector field as
in this paper. A
realisation of such a scenario can be found in Refs. \cite{cogollo,valenzuela} (see also Ref. \cite{leblond}).} %For examples where the linear terms
%dominate see ref. \cite{lv} and where quadratic terms dominate see refs. \cite{lr1,ev}.}:
\begin{itemize}
\item Vector field spectrum ($\calp_{\zeta_A}$) and bispectrum ($\calb_{\zeta_A}$) dominated
by the tree-level terms \cite{dkl}.
\item Vector field spectrum ($\calp_{\zeta_A}$) and bispectrum ($\calb_{\zeta_A}$) dominated
by the 1-loop contributions.
\item Vector field spectrum ($\calp_{\zeta_A}$) dominated by the tree-level terms and
bispectrum ($\calb_{\zeta_A}$) dominated by the 1-loop contributions.
\item Vector field spectrum ($\calp_{\zeta_A}$) dominated by the 1-loop contributions and
bispectrum ($\calb_{\zeta_A}$) dominated by the tree-level terms.
\end{itemize}
In order to study these possibilities, we first need to estimate the integrals coming from loop
contributions. From Eqs. (\ref{Pzetal}), (\ref{bsl}), (\ref{ssl1}), and (\ref{bssl1}) the integrals to solve are:
\bea
\calpz^{\rm 1-loop}(\bfk)&=& \frac{1}{2} N_{AA}^{ij} N_{AA}^{kl} \int \frac{d^3p k^3}{4\pi p^3|\bfk + \bfp|^3}
\calt_{ik}(\bfk+\bfp)\calt_{jl}(\bfp) \,, \label{intsl}\\
\calbz^{\rm 1-loop}(\bfk,\bfk',\bfk'')&=&N_{AA}^{ij}N_{AA}^{kl}N_{AA}^{mn}\int\frac{d^3pk^3k'^3}{4\pi p^3|
\bfk+\bfp|^3|\bfk'-\bfp|^3}\calt_{il}(\bfp)\calt_{kn}(\bfk+\bfp)\calt_{jm}(\bfk'-\bfp) \,. \label{intbsl}
\eea
The above integrals cannot be done analytically, but they can be estimated in the same way as that presented in 
Refs.\cite{lyth0,lyth2,lyth1}: these papers show that the integrals are proportional to $\ln(kL)$
where $L$ is the box size. To evaluate them we take the spectrum to be scale-invariant, which will be a good approximation if both scalar
field $\phi$ and vector field  ${\bf A}$ are sufficiently light during inflation. The integrals are logarithmically
divergent at the zeros of the denominator and in each direction, but there is a cutoff at $k\sim L^{-1}$. We found
that in our case the integrals are also proportional to $\ln(kL)$ and that each singularity gives equal
contributions to the overall result.  We find from Eqs. (\ref{intsl}) and (\ref{intbsl}):
\bea
\calp_{\zeta A}^{\rm 1-loop} (\bfk)&=&\frac{1}{2}N_{AA}^{ij}N_{AA}^{kl}(2\calp_++\calp_{\rm long})\delta_{ik}\calt_{jl}
(\bfk)\ln(kL) \,, \label{sploop}\\
\calb_\zeta^{\rm 1-loop}(\bfk,\bfk',\bfk'')&=&N_{AA}^{ij}N_{AA}^{kl}N_{AA}^{mn}\ln(kL)\big(2\calp_++\calp_{\rm long})\delta_{il}\big[\calt_{kn}(\bfk)\calt_{jm}
(\bfk')\big] \,. \label{bsploop}
\eea
Except when considering low CMB multipoles, the box size should be set at $L\simeq H_0$ \footnote{
$H_0$ is the Hubble parameter today.} \cite{leblond,klv}, 
giving $\ln(kL)\sim 1$ for relevant cosmological scales. 

\section{Estimating $f_{NL}$}
The non-gaussianity parameter is defined by \cite{ks,maldacena}\footnote{We employ the WMAP
sign convention.}:
\be
\fnl=\frac{5}{6}\frac{\calbz(\bfk,\bfk',\bfk'')}{\big[\calpz(k)\calpz(k')+{\rm cyc.\  perm.}\big]}\label{fnl1} \,.
\ee
Since the isotropic contribution to the curvature perturbation 
is always dominant compared to the anisotropic one, we can write in the above expression only the 
isotropic part of the spectrum $\calpz^{\rm iso}(k)$:
\be
\fnl=\frac{5}{6}\frac{\calbz(\bfk,\bfk',\bfk'')}{\big[\calpz^{\rm iso}(k)\calpz^{\rm iso}(k')+{\rm cyc.\  perm.}\big]} \,.
\label{fnl2}
\ee
Keeping in mind the above expression, we will estimate the possible amount of non-gaussianity generated by the 
anisotropic part of the primordial curvature perturbation. To do it we take into account the different possibilities 
mentioned in the previous section, where the non-gaussianity is produced solely by vector field perturbations.
 
\subsection{Vector field spectrum ($\calp_{\zeta_A}$) and bispectrum ($\calb_{\zeta_A}$)
dominated by the tree-level terms} 
We start our analysis by considering the case studied in Ref. \cite{dkl}, where the authors assume that
the bispectrum is dominated by vector fields perturbations and that the higher order
contributions from the vector field are always subdominant, i.e., $N_A^i\delta A_i\gg N_{AA}^{ij}\delta A_i \delta A_j$. 
This means that both the spectrum and the bispectrum are dominated by the tree-level terms, i.e., $\calpz_A^{\rm tree} \gg \calpz_A^{\rm 1-loop}$ and $\calbz_A^{\rm tree} \gg \calbz_A^{\rm 1-loop}$, so that
the level of non-gaussianity $f_{NL}$ is given by:
\be
\fnl\:=\: \frac{5}{6} \frac{\calb_\zeta^{\rm tree} (\bfk,\bfk',\bfk'')}{\big[\calpz^{\rm iso}(k)\calpz^{\rm iso}(k')+{\rm cyc.\  perm.}\big]}\:\simeq\:\frac{5}{6}\frac{\calbz_A^{\rm tree} (\bfk,\bfk',\bfk'')}{\big[\calpz^{\rm iso}(k)\calpz^{\rm iso}(k')+{\rm cyc.\  perm.}\big]} \,.
\label{fnlm1}
\ee
Since the anisotropic contribution to the curvature perturbation is subdominant, we can take
$\calpz\sim\calpz^{\rm iso}$, so we may write:
\be
\fnl\:\simeq\:\frac{N_A^i N_A^k N_{AA}^{mn}\big[\calt_{im}(\bfk)\calt_{kn}(\bfk')+{\rm cyc.\  perm.}\big]}{\big[\calpz(k)\calpz(k')+{\rm cyc.\  perm.}
\big]} \,.
\label{fnlm2}
\ee
Assuming that $\calp_{\rm long}$, $\calp_+$, and $\calp_-$ are all of the same order of
magnitude, and that the spectrum is scale invariant, we
may write the above equation as:
\be
\fnl\:\simeq\:\frac{\calp_A^2 N_A^2 N_{AA}}{\calpz^2} \,,
\label{fnlm3}
\ee
where $\calp_A=2\calp_++\calp_{\rm long}$. Taking as a typical value for the vector field perturbation $\delta 
A=\sqrt{\calp_A}$ and $ N_A\delta A > N_{AA}\delta A^2$, the contribution of the vector field to $\zeta$  is given by 
$\zeta_A\sim \sqrt{\calpz_A}\sim N_A \sqrt{\calp_A}$. %where $\calpz_A$ is the spectrum of the anisotropic curvature 
%perturbation.
Thus, we may write an upper bound for $\fnl$:
\be
\fnl\:\lsim\:\frac{\calpz_A\threehalf}{\calpz^2} \,.
\label{fnlm4}
\ee
Since the level of statistical anisotropy in the power spectrum is of order $\gz\sim\calpz_A/\calpz$,
 and since $\calpz\half \simeq 5\times 10 ^{-5}$ \cite{wmap5}, \eq{fnlm3} yields \cite{dkl}:
\be
\fnl\:\lsim\:10^3\(\frac{\gz}{0.1}\)\threehalf \,.
\label{fnlm5}
\ee
The above expression gives an upper bound for the level of non-gaussianity $\fnl$ in terms of
the level of statistical anisotropy in the power spectrum $\gz$ when the former is generated
by the anisotropic contribution to the curvature perturbation. As we may see, the current
observational limit on $f_{NL}$, $f_{NL} < 111$ \cite{wmap5}, may easily be exceeded.

As an example of this model, we apply the previous results to a specific model, e.g. the vector curvaton 
scenario \cite{vc,vc2,RA2}, where the $N$-derivatives are \cite{dkl}:
\bea
N_{A}&=&\frac{2}{3A}r \,,\label{navc}\\
N_{AA}&=&\frac{2}{A^2}r\label{naavc} \,,
\eea
where $A\equiv|\bfA|$ is the value of vector field just before the vector curvaton field decays
and the parameter $r$ is the ratio between the energy density of the vector curvaton field and
the total energy density of the Universe just
before the vector curvaton decay. We begin exploring the conditions under which the vector field 
spectrum and bispectrum are always dominated by the tree-level terms. From Eqs. (\ref{Pzetat}), (\ref{bzt}), 
(\ref{sploop}) and (\ref{bsploop}) our constraint leads to:
\bea
\calp_A N_A^2 &\gg & \calp_A^2 N_{AA}^2 \,,\\
\calp_A^2 N_A^2 N_{AA}&\gg &\calp_A^3 N_{AA}^3 \,.
\eea
Thus, it follows that:
\be
\calp_A \ll \(\frac{N_A}{N_{AA}}\)^2 \,.
\ee
We have to remember that in the present case the contribution of the vector field to $\zeta$  is given by 
$\zeta_A\sim \sqrt{\calpz_A}\sim N_A \sqrt{\calp_A}$. Then, the above equation combined with Eqs. (\ref{navc}) and 
(\ref{naavc}) leads to:
\be
r \gg 2.25\times 10^{-4}\gz\half \,.\label{rbt}
\ee
This is a lower bound on the $r$ parameter we have to consider when building a realistic
particle physics model of the vector curvaton scenario.

Finally, from \eq{fnlm3}, the $f_{NL}$ parameter in this scenario is given by:
\be
\fnl\simeq \frac{4.5\times 10^{-2}}{r}\(\frac{\gz}{0.1}\)^2.
\ee
This is a consistency relation between $f_{NL}$, $g_\zeta$, and $r$ which will help when
confronting the specific vector curvaton realisation against observation.

\subsection{Vector field spectrum ($\calp_{\zeta_A}$) and bispectrum ($\calb_{\zeta_A}$)
dominated by the 1-loop contributions}
Since the bispectrum is dominated by 1-{\rm loop} contributions and is given by \eq{bsploop}, we may 
write \eq{fnl2} as:
\be
\fnl\:\simeq\:\frac{N_{AA}^{ij}N_{AA}^{kl}N_{AA}^{mn}\ln(kL)\big(2\calp_++\calp_{\rm long})\delta_{il}\big[\calt_{kn}
(\bfk)\calt_{jm}(\bfk')+{\rm cyc.\  perm.}\big]}{\big[\calpz(k)\calpz(k')+{\rm cyc.\  perm.}\big]} \,.
\label{fnl11}
\ee
Assuming again that $\calp_{\rm long}$, $\calp_+$, and $\calp_-$ are all of the same order of
magnitude, and that the spectrum is scale invariant, the above equation leads to:
\be
\fnl\:\simeq\:\frac{\calp_A^3 N_{AA}^3}{\calpz^2} \,.
\label{fnl12}
\ee
Since the vector field spectrum is dominated by the 1-{\rm loop} contribution, 
$\zeta_A\sim \sqrt{\calpz_A}\sim N_{AA} 
\calp_A$. Thus, and taking into account that $\gz\sim\calpz_A/\calpz$ and $\calpz\half\simeq 
5\times 10 ^{-5}$ \cite{wmap5}, we 
find:
\be
\fnl\:\sim\:\frac{1}{\sqrt{\calpz}}\(\frac{\calpz_A}{\calpz}\)\threehalf\:\sim\:10^3\(\frac{\gz}{0.1}\)\threehalf.
\label{fnl13}
\ee
The biggest difference between the result found in Ref. \cite{dkl}, given by \eq{fnlm5}, and the result given by 
\eq{fnl13}, is that the latter gives an equality relation between the non-gaussianity parameter $\fnl$ and the 
level of statistical anisotropy in the power spectrum $\gz$. Following the recent bounds for
$\fnl$: $-9 < f_{NL} <111$ 
\cite{wmap5}, this scenario predicts an upper bound for the $\gz$ parameter:
\be
\gz<\,0.02 \,.\label{gz}
\ee
This bound is stronger than that obtaining from direct observations in Ref. \cite{ge2}.

Again we apply our result to the vector curvaton scenario. Since we are assuming that the vector field 
spectrum and bispectrum are dominated by 1-loop contributions, we get from Eqs. (\ref{Pzetat}), (\ref{bzt}), 
(\ref{sploop}), and (\ref{bsploop}):
\be
\calp_A > \(\frac{N_A}{N_{AA}}\)^2 \,,
\ee
which for the vector curvaton scenario becomes:
\be
r <2.25\times 10^{-4}\gz\half \,. \label{rbl}
\ee
%and $\fnl$ is given by \eq{fnl13}.
This is a lower bound on the $r$ parameter we have to consider when building a realistic
particle physics model of the vector curvaton scenario.

\subsection{Vector field spectrum ($\calp_{\zeta_A}$) dominated by the tree-level terms and
bispectrum ($\calb_{\zeta_A}$) dominated by the 1-loop contributions}
In order to check the viability of this case, we start studying the implications of the restrictions over the 
spectrum and the bispectrum, i.e., what happens when we assume that the vector field spectrum
is dominated by the tree-level terms and the bispectrum is dominated by the 1-loop contributions.
From Eqs. (\ref{Pzetat}), (\ref{bzt}), 
(\ref{sploop}), and (\ref{bsploop}) it follows that:
\bea
\calp_A N_A^2 \gg  \calp_A^2 N_{AA}^2 &\Rightarrow & \calp_A\ll\frac{N_A^2}{N_{AA}^2} \,, 
\label {fc}\\
\calp_A^2 N_A^2 N_{AA}\ll  \calp_A^3 N_{AA}^3 &\Rightarrow & \calp_A\gg\frac{N_A^2}{N_{AA}^2} \,.
\label{sc}
\eea 
As we may see, it is impossible to satisfy simultaneously Eqs. (\ref{fc}) and (\ref{sc}).
This is perhaps related to the fact that we have
taken into account only one vector field. Such a conclusion may be relaxed if we take into
account more than one vector field, as analogously happens in the scalar multifield case
\cite{cogollo,valenzuela}.

\subsection{Vector field spectrum ($\calp_{\zeta_A}$) dominated by the 1-loop contributions
and bispectrum ($\calb_{\zeta_A}$) dominated by the tree-level terms}
As in the previous case, it is impossible to satisfy the conditions under which
the spectrum is always dominated by the 1-loop contributions and the bispectrum is always dominated
by the tree-level terms: 
%we arrived to an inconsistency. In this particular case the conditions also are:
\bea
\calp_A &\gg& \frac{N_A^2}{N_{AA}^2} \,, \\
\calp_A &\ll& \frac{N_A^2}{N_{AA}^2} \,.
\eea
%it the bispectrum is to be dominated by the tree-level terms. 
Again, the conclusion may be relaxed if we take into account more than one vector field.

\section{Conclusions}
We have studied in this paper the order of magnitude of the level of non-gaussianity in the
bispectrum $f_{NL}$ when statistical anisotropy is generated by the presence of one vector
field. Particularly, we have shown that it is possible to get a high level of non-gaussianity
if we assume that the 1-loop contributions dominate
over the tree-level terms in both the vector field spectrum ($\calp_{\zeta_A}$) and the
bispectrum ($\calb_{\zeta_A}$). $\fnl$ is
given in this case by \eq{fnl13}, where we may see that there is a consistency relation
between $\fnl$ and the amount 
of statistical anisotropy in the spectrum $\gz$. Such a consistency relation lets us fix one
of the two parameters, i.e., if the non-gaussianity in the bispectrum
is detected and our scenario is appropriate, the amount of statistical anisotropy in the
power spectrum must have a specific value, which is given 
by \eq{fnl13}. A similar conclusion is reached if the statistical anisotropy in the power
spectrum is detected before the non-gaussianity in the bispectrum is. As an example we may see the 
result given in \eq{gz}, where an indirect (but stronger than the observational) upper bound
on $\gz$ is obtaining from the current upper observational bound on $\fnl$.

\section*{Acknowledgments}
This work was   supported by STFC  grant ST/G000549/1
and by  EU grant MRTN-CT-2006-035863. C.A.V.-T. acknowledges the Department of Physics at Lancaster University in Lancaster (UK) were part 
of this work was done.

%%%%%%%%%%%%%%%%%%%%%%%%%%%%%%%%%%%%%%%%%%%%%%%%%%%%%%%%%%%%%%%%%%%%%%%%%%%%%%%%%%%%%%%%%%%%%%%%%%
%%%%%%%%%%%%%%%%%%%%%%%%%%%%%%%%%%%%%%%%%%%%%%%%%%%%%%%%%%%%%%%%%%%%%%%%%%%%%%%%%%%%%%%%%%%%%%%%%%
%%%%%%%%%%%%%%%%%%%%%%%%%%%%%%%%%%%%%%%%%%%%%%%%%%%%%%%%%%%%%%%%%%%%%%%%%%%%%%%%%%%%%%%%%%%%%%%%%%
%%%%%%%%%%%%%%%%%%%%%%%%%%%%%%%%%%%%%%%%%%%%%%%%%%%%%%%%%%%%%%%%%%%%%%%%%%%%%%%%%%%%%%%%%%%%%%%%%%

%\newpage

%%%%%%%%%%%%%%%%%%%%%%%%%%%%%%%%%%%%%%%%%%%%%%

\renewcommand{\refname}{{\large References}}

\end{document}